\documentclass[superscriptaddress,aps,prl,twocolumn,floatfix]{revtex4-2}

\usepackage{graphicx}
\usepackage{dcolumn}
\usepackage{bm}
\usepackage{multirow}%
\usepackage{amsmath,amssymb,amsfonts}%
\usepackage{amsthm}%
\usepackage{mathrsfs}%
\usepackage{bbm}%
\usepackage{stackengine}
 \usepackage[title]{appendix}%
 \usepackage{xcolor}%
 \usepackage{textcomp}%
 \usepackage{booktabs}%
 \usepackage{algorithm}%
 \usepackage{algorithmicx}%
 \usepackage{algpseudocode}%
 \usepackage{listings}%
\usepackage{hyperref}%
\usepackage{xurl}
\usepackage{color}
\definecolor{mygreen}{rgb}{0,0.5,0}
\definecolor{mygrey}{rgb}{0.5,0.5,0.5}
\definecolor{myred}{rgb}{0.75,0,0}
\definecolor{myblue}{rgb}{0,0,0.75}
\definecolor{mymagenta}{cmyk}{0,1,0,0.12}
\definecolor{mycyan}{cmyk}{1,0,0,0.12}
\definecolor{myorange}{rgb}{1.,0.5,0}
\definecolor{myviolet}{rgb}{0.6,0.15,0.6}
\definecolor{mybrown}{cmyk}{0,0.50,1,0.41}
\usepackage[normalem]{ulem}

\newcommand{\gtext}[1]{{\color{mygreen}#1}}

\newcommand{\bF}{\mathbf{F}}

\newcommand{\bN}{\mathbf{N}}

\newcommand{\bFmax}{{\mathbf{F}_\mathrm{max}}}

\newcommand{\subdc}{_\mathrm{dc}}
\newcommand{\subrf}{_\mathrm{rf}}

\newcommand{\bvdc}{\mathbf{B}\subdc}

\newcommand{\bvrf}{\mathbf{B}\subrf}

\newcommand{\subOP}{_\mathrm{OP}}
\newcommand{\subat}{_\mathrm{at}}
\newcommand{\dc}{dc}

\usepackage{siunitx}
\DeclareSIUnit\torr{Torr}
\DeclareSIUnit\amagat{amg}
\newcommand{\SM}{End Matter}

\begin{document}

\title[Article Title]{Quantum noise scaling in continuously operating multiparameter sensors}

\newcommand{\ICFO}{ICFO - Institut de Ci\`encies Fot\`oniques, The Barcelona Institute of Science and Technology, 08860 Castelldefels (Barcelona), Spain}
\newcommand{\ICREA}{ICREA - Instituci\'{o} Catalana de Recerca i Estudis Avan{\c{c}}ats, 08010 Barcelona, Spain}

\author{Aleksandra Sierant}
\email{aleksandra.sierant@icfo.eu}
\affiliation{\ICFO}

 \author{Diana Méndez-Avalos}
 \affiliation{\ICFO}

 \author{Santiago Tabares Giraldo}
 \affiliation{\ICFO}

\author{Morgan W. Mitchell}
\affiliation{\ICFO}
\affiliation{\ICREA}

\begin{abstract}
We experimentally investigate the quantum noise mechanisms that limit continuously operating multiparameter quantum sensors. Using a hybrid rf–dc optically pumped magnetometer, we map the photon shot noise, spin projection noise, and measurement back-action noise over an order of magnitude in probe power and a factor of three in pump power while remaining quantum-noise-limited.  We observe linear, quadratic, and cubic scaling of the respective total noise powers with probe photon flux, together with a quadratic dependence of back-action on pump photon flux, in quantitative agreement with a stochastic Bloch-equation model. At higher probe powers, additional probe-induced relaxation modifies the spin-noise spectrum while preserving the integrated noise scaling. Our results reveal fundamental, resource-dependent trade-offs unique to continuously monitored multiparameter sensors and establish experimentally the quantum limits governing their optimal operation. 
\end{abstract}

\maketitle

Paradigmatic quantum sensors, including magnetometers \cite{Kominis2003}, gyroscopes \cite{KornackPRL2005, WalkerBook2016}, and instruments to search for physics beyond the standard model \cite{HunterS2013, Afach2021N}, employ continuous, non-destructive optical monitoring of a spin ensemble as it evolves in response to environmental fields and inertial effects. A variety of quantum noise sources, including photon shot noise (PSN), spin-projection noise (SPN) and measurement back-action noise (MBA) contribute to the intrinsic quantum noise of these sensors, and thus to their quantum-noise-limited sensitivity \cite{Helstrom1969, Degen2017, PezzeRMP2018}. These several contributions scale differently with the employed particle-number resources, i.e., with the number of spins and with the optical powers used to probe and to optically pump the spin ensemble. In the best-known example, first studied in the context of gravitational-wave detection by Braginsky \cite{BraginskyBook1995}, the PSN contribution decreases with increasing probe flux, whereas the MBA contribution increases. This creates, \textit{ceteris paribus}, an optimum probe power or measurement strength, and a quantum limit to the sensitivity. 

More broadly, scaling of the different quantum noise contributions with the three resource inputs defines a complex optimization landscape that determines the ultimate potential of such sensors  \cite{MitchellQST2017}. Important variants of this problem allow for non-classical, e.g., squeezed, states \cite{GanapathyPRX2023} and for multi-parameter sensing \cite{LiS2026}, in which more than one influence on the spins is simultaneously detected. Prior work has studied the sensitivity landscape of spin-noise spectroscopy  (for which there is no pump and little sensitivity to external fields), with atom number and probe flux, with and without squeezed probes \cite{Lucivero2016, Lucivero2017a}. Scaling of SPN and PSN with atom number has been studied in a single-parameter magnetometer \cite{TroullinouPRL2023}, whereas MBA could not be studied in that system due to MBA-evasion \cite{Troullinou2021}. Here we complete this tableau by studying scaling of all three contributions with probe and pump flux, with and without a squeezed probe. As described in \cite{SqhOPM2025} it is natural to study this in the multi-parameter sensing scenario \cite{lipka2024multiparameter}, which, in contrast to the single-parameter scenario, cannot be MBA-evading.

Specifically, we experimentally map the scaling of these quantum noise contributions in a hybrid rf–dc optically pumped magnetometer (hOPM) \cite{lipka2024multiparameter, SqhOPM2025}, a multi-parameter sensor that measures magnetic field in two separated frequency bands with a single spin ensemble. 
We independently vary the probe and pump powers that control measurement strength and spin polarization. We observe that PSN, SPN, and MBA exhibit distinct linear, quadratic, and cubic scaling with probe power, respectively, while MBA additionally displays a quadratic dependence on pump power. These results reveal fundamental, resource-dependent trade-offs in continuously monitored single- and multi-parameter sensors, and establish experimentally the quantum noise behaviors governing their optimal operation.

\begin{figure*}[t]
    \centering
    \includegraphics[width=1\textwidth]{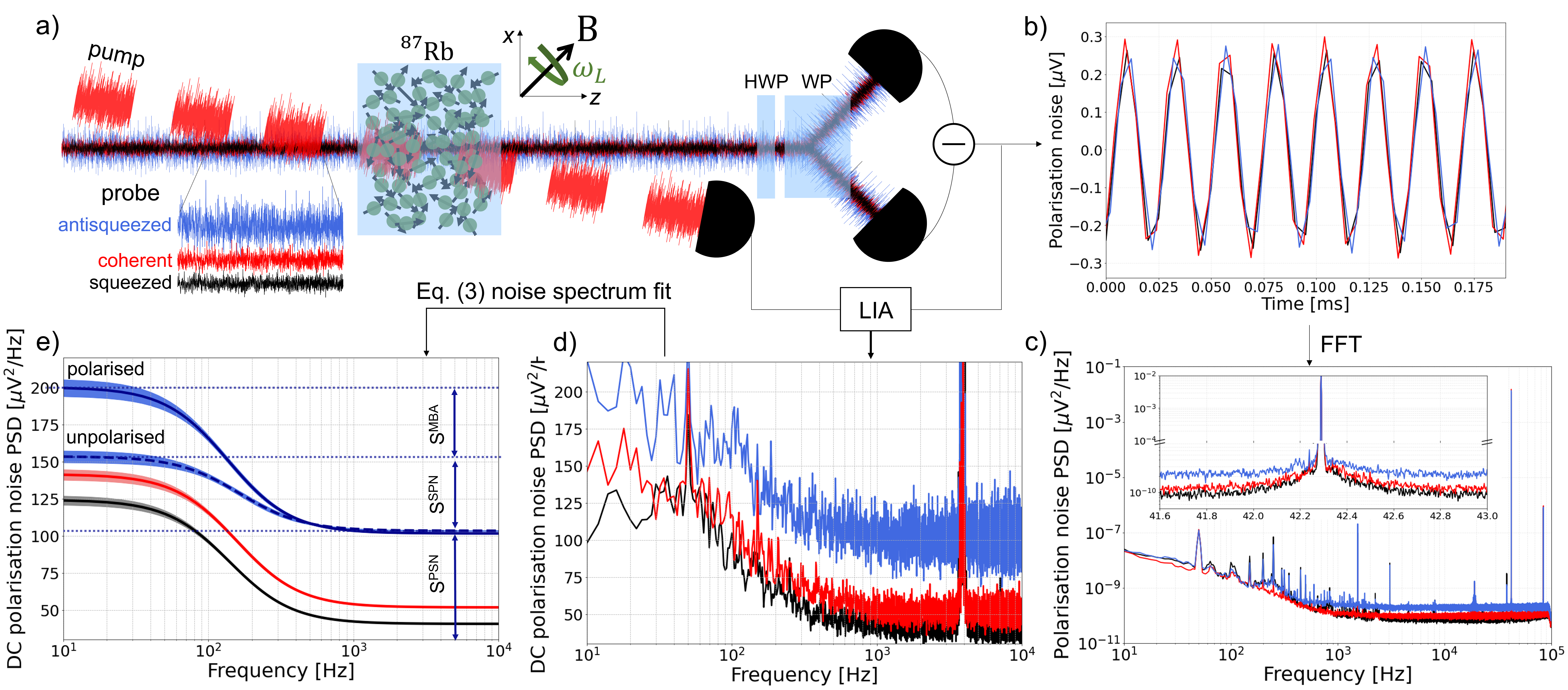}
\caption{\textbf{Quantum noise characterization scheme and methodology.}
(a) Schematic of the experimental setup. $^{87}$Rb vapor is optically pumped in the Bell--Bloom configuration. A dc magnetic field $\mathbf{B}_{\dc}$ applied in the $x$--$z$ plane at $45^{\circ}$ with respect to the pump--probe direction enables hybrid dc/rf magnetometry \cite{LipkaPRAppl2024}. The polarization rotation of the probe beam, prepared in a squeezed (black), coherent (red), or antisqueezed (blue) state, is detected using a Wollaston prism (WP) and a balanced photodetector. HWP - half wave plate, LIA - lock-in amplification.
(b) Representative time-domain polarization-rotation signals for the three probe states.
(c) Measured polarization noise spectra with a zoomed inset around the Larmor frequency for the three probe states, polarized atoms. PSD - power spectral density. Probe power \SI{2.8}{m\watt}, pump power  \SI{10}{\micro\watt}.
(d) Demodulated polarization noise PSD of, obtained by a digital lock-in with the pump power as phase reference \cite{LipkaPRAppl2024}. 
(e) Fits to the demodulated noise spectra for all three probe states. PSN, SPN and MBA are indicated only for the antisqueezed probe state, for clarity. Shaded regions indicate $\pm1\sigma$ uncertainty obtained by bootstrapping. }
    \label{fig:setup}
\end{figure*}

\newcommand{\PRLsection}[1]{\noindent \textit{#1} ---} 
\newcommand{\subpr}{_\mathrm{pr}}
\newcommand{\subpu}{_\mathrm{pu}}
\newcommand{\submax}{_\mathrm{max}}

\PRLsection{Experimental system and noise analysis}
The experimental scheme and noise-analysis methodology are illustrated in Fig.~\ref{fig:setup}; further details are provided in the \SM. A shielded $^{87}$Rb vapor cell is optically pumped in the Bell--Bloom configuration, producing a collective spin polarization that precesses in an applied dc magnetic field oriented at $45^\circ$ with respect to the pump--probe axis. The collective atomic spin $\mathbf{F}$ evolves under the influence of a magnetic field and continuous optical measurement. To leading order, the spin dynamics are described by the stochastic Bloch equation
\begin{eqnarray}
\label{eq:SpinDynamics}
\frac{d}{dt}\bF(t)  &=&  \left[-\gamma \mathbf{B}(t) + G_S S_3(t) \hat{z} \right] \times \bF(t)
- (\Gamma+\Gamma\subpr)\bF(t)
\nonumber \\ & & + R\subOP(t)\,[\bFmax -\bF(t)] +  \bN\subat(t) + \bN\subpr(t)
\end{eqnarray}
where $\gamma$ is the atomic gyromagnetic ratio, $\mathbf{B}$ is the magnetic field vector, $\Gamma$ is the intrinsic spin relaxation rate, $\Gamma\subpr=\alpha P\subpr$ is the probe-induced relaxation rate proportional to probe power $P\subpr$, and $R\subOP$ is the optical pumping rate. The vector $\bFmax$, proportional to the number of atoms in the ensemble, denotes the fully polarized spin state. The Langevin noise terms $\bN\subat(t)$ and $\bN\subpr(t)$ describe intrinsic spin fluctuations and probe-induced relaxation noise, respectively. The term $G_S S_3(t)\hat{z}$ represents optical Zeeman shifts induced by fluctuations of the probe ellipticity and gives rise to measurement back-action. 

The resulting polarization rotation (Fig.~\ref{fig:setup}b,c) of an off-resonant probe beam is detected with a balanced polarimeter by the readout of $S_2(t)$ Stokes component
\begin{equation}
\label{eq:MeasurementModel}
S_2^{\mathrm{(out)}}(t) = G_F  F_z(t) S_1^{\mathrm{(in)}}(t) + N_{\mathrm{opt}}(t),
\end{equation}
where $\phi = G_F F_z$ is the polarization rotation angle, $G_F$ is a detuning-dependent coupling factor, and $N_{\mathrm{opt}}$ denotes the quantum polarization noise of the detected Stokes component, PSN \cite{PhysRevD.23.1693}. The approximation holds for small rotation angles. Together, the spin dynamics and optical readout define a unified framework for PSN, SPN and MBA in a continuously operating multiparameter quantum sensor. The probe is prepared in coherent, squeezed, or antisqueezed polarization states, enabling controlled redistribution of optical quantum noise between the conjugate Stokes components $S_2$ and $S_3$. Due to optical losses in the probe path and detection system, the measured levels of squeezing and antisqueezing are asymmetric, with antisqueezing exceeding the squeezing reduction, as expected for squeezed states after optical losses.  The pump power is varied between 5 and $15~\mu$W, and the probe power between 0.5 and $3~$mW.

The detected signal is demodulated at the pump frequency to obtain polarization noise spectra (Fig.~\ref{fig:setup}d). The measured polarization noise inherits the dynamical response of the collective spin. As a result, the quantum noise spectrum of the demodulated signal consists of a frequency-independent PSN contribution and atomic noise contributions filtered by the magnetic resonance. The single-sided noise power spectral density of the measured quadratures can be written as
\begin{equation}
\label{eq:NoiseSpectrum}
\mathcal{S}_{\mathrm{dc},\mathrm{rf}}(f)
=
\xi^{2}\,\mathcal{S}^{\mathrm{PSN}}
+
\mathcal{L}(f)
\left(
\mathcal{S}^{\mathrm{SPN}}
+
\bar{\xi}^{2}\,\mathcal{S}^{\mathrm{MBA}}
\right),
\end{equation}
where $\xi^{2}$ ($\bar{\xi}^{2}$) denotes the squeezing (antisqueezing) factor of the detected Stokes component, and $\mathcal{L}(f)=\Delta f^{2}/(f^{2}+\Delta f^{2})$ is the Lorentzian response associated with the magnetic-resonance linewidth $\Delta f$, which may differ slightly for unpolarised and polarised ensembles due to additional relaxation channels. Photon shot noise $\mathcal{S}^{\mathrm{PSN}}$ is spectrally flat, while spin projection noise $\mathcal{S}^{\mathrm{SPN}}$ and measurement back-action noise $\mathcal{S}^{\mathrm{MBA}}$ are shaped by the Lorentzian envelope. Polarization squeezing redistributes quantum fluctuations between conjugate Stokes components, reducing the detected shot noise while simultaneously enhancing the back-action contribution, thereby preserving the underlying spectral structure but modifying the relative noise weights.

Narrow technical noise peaks visible around \SI{50}{\hertz} and \SI{4}{\kilo\hertz} were excluded from the fits. Model parameters were obtained using maximum-likelihood estimation combined with bootstrapping, following Ref.~\cite{SqhOPM2025}. 
For each test condition we fit the demodulated spectra with \autoref{eq:NoiseSpectrum}, as illustrated in Fig.~\ref{fig:setup}d, to identify the noise contributions $\mathcal{S}$. For SPN and MBA, we compute the total power as $\mathcal{S}_{\mathrm{tot}} = \mathcal{S}\,\Delta f \pi/2$, which is the integral of the corresponding noise PSD over positive frequencies. The SPN contribution was extracted from measurements on an unpolarized ensemble by subtracting the spectrally flat PSN term $\xi^{2}\mathcal{S}^{\mathrm{PSN}}$ in Eq.~(\ref{eq:NoiseSpectrum}) from the fitted noise spectrum. The MBA contribution was isolated from measurements on a polarized ensemble by subtracting both the PSN term and the SPN contribution inferred from the unpolarized ensemble. The inferred noise levels identify the frequency regimes in which the hOPM is limited by SPN and MBA at low frequencies and by PSN at high frequencies.

\begin{figure*}[t]
    \centering
    \includegraphics[width=1\textwidth]{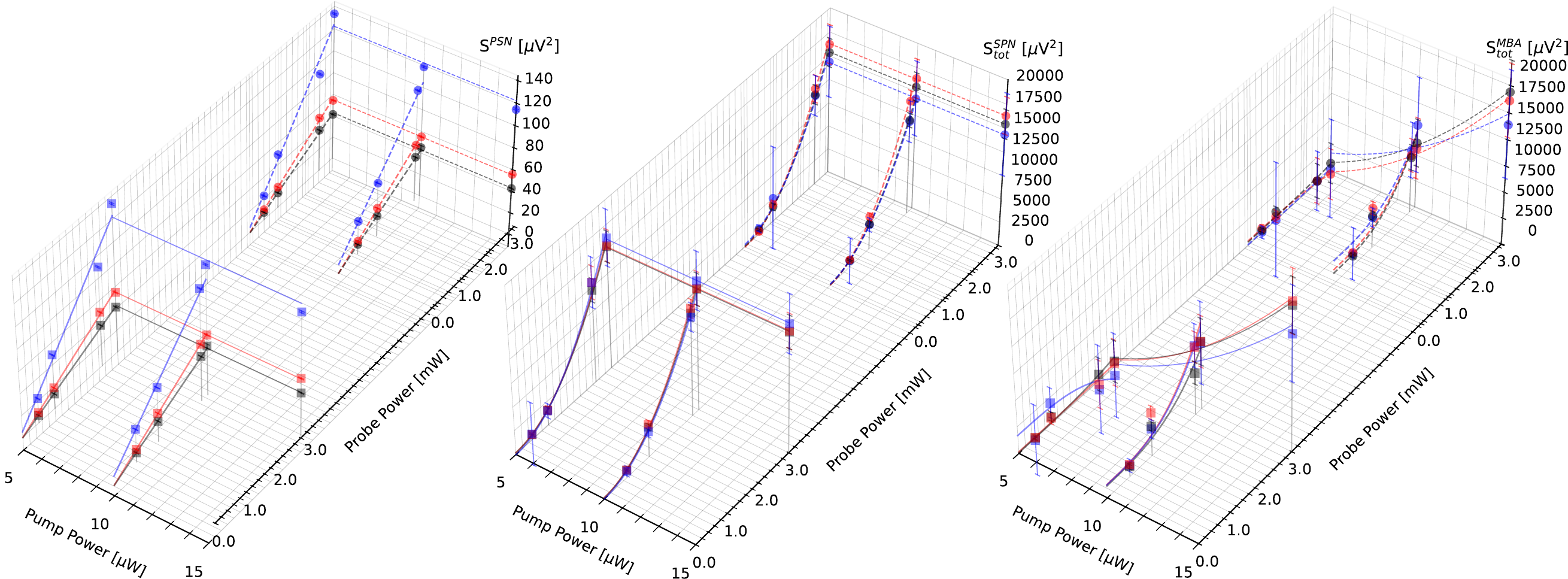}
\caption{
\textbf{Quantum noise scaling with probe and pump power.}
Measurements for the dc (squares, solid fits) and rf (circles, dashed fits) channels using coherent (red), squeezed (black), and antisqueezed (blue) probe light. Error bars show $\pm 1 \sigma$ uncertainty obtained by bootstrapping.
\textbf{Left:} PSN power spectral density $\mathcal{S}^{\mathrm{PSN}}$, exhibiting linear scaling with probe power, $\mathcal{S}^{\mathrm{PSN}} = a_0 + a_1 P_{\mathrm{pr}}$, and no dependence on pump power.
\textbf{Middle:} Total SPN power for an unpolarized ensemble, showing quadratic scaling with probe power, $\mathcal{S}^{\mathrm{SPN}}_{\mathrm{tot}} = a_0 + a_2 P_{\mathrm{pr}}^{2}$, and no dependence on pump power.
\textbf{Right:} Total MBA noise power for polarized ensembles, displaying cubic scaling with probe power, $\mathcal{S}^{\mathrm{MBA}}_{\mathrm{tot}} = a_0 + a_3 P_{\mathrm{pr}}^{3}$, and a quadratic dependence on pump power.
Fit parameters for all panels are given in \SM.
}
    \label{fig:allNoises}
\end{figure*}

\PRLsection{Quantum noise scaling} 
The stochastic Bloch equation Eq.~(\ref{eq:SpinDynamics}) and optical readout Eq.~(\ref{eq:MeasurementModel}) predict distinct scaling laws for the fundamental quantum noise contributions as a function of the optical powers. PSN, arising from quantum fluctuations of the probe polarization, scales linearly with probe power, $\mathcal{S}^{\mathrm{PSN}}\propto P\subpr$, and is independent of the pump power. SPN, originating from intrinsic quantum fluctuations of the collective spin, is transduced to the optical readout with a strength proportional to the probe intensity, leading to a quadratic scaling $\mathcal{S}^{\mathrm{SPN}}\propto P\subpr^{2}$ at low probe powers. At higher probe powers, off-resonant scattering from the probe light induces significant probe-induced relaxation, which we model by an effective relaxation rate $\Gamma\subpr=\alpha P\subpr$. This term broadens the spin-noise spectrum while preserving the total spin-noise power $\mathcal{S}^{\mathrm{SPN}}_{tot} \propto P\subpr^{2}$. MBA arises from fluctuations of the probe ellipticity, described by the $S_3$ Stokes component, which act on the spin through optical Zeeman shifts. The  back-action torque $G_S S_3(t) \hat{z} \times \mathbf{F}$ contains the product of a quantum fluctuation $S_3(t)$, with noise power proportional to the probe intensity, and the spin $\mathbf{F}$, with mean value proportional to pump power. The MBA contribution to the variance of $\mathbf{F}$ thus scales as $P\subpr P_{\mathrm{pu}}^2$. The readout process converts this to observed signal with a gain proportional to $P\subpr^2$.  The model thus predicts total noise power $\mathcal{S}^{\mathrm{MBA}}_{tot} \propto P\subpr^{3} P_{\mathrm{pu}}^2$, i.e., MBA power scaling cubically with probe power, and quadratically with pump power. This scaling law enables an experimental test of the quantum noise mechanisms governing continuously operating multiparameter sensors, independently of the specific details of the sensor. Figure~\ref{fig:allNoises} summarizes the measured PSN, SPN, and MBA contributions to the polarization noise as functions of the probe and pump oprical power for the dc and rf readout channel. 

\gtext{}

\PRLsection{PSN} For all probe quantum states the PSN exhibits a linear dependence on the probe power and no measurable dependence on the pump power, consistent with the expected scaling $\mathcal{S}^{\mathrm{PSN}}\propto P\subpr$. Linear fits of the form $\mathcal{S}^{\mathrm{PSN}} = a_1 P\subpr + a_0$ are shown in Fig.~\ref{fig:allNoises}(left) and fit details are summarized in the \SM. The extracted slopes $a_1$ are independent of pump power within uncertainties and are nearly identical for polarized and unpolarized ensembles, confirming that PSN is set solely by the probe field. The slopes follow the ordering imposed by the probe quantum state: compared to coherent probing, PSN is reduced for squeezed light and increased for antisqueezed light, reflecting the redistribution of quantum fluctuations between the detected Stokes component and its conjugate. The relative noise reduction and enhancement are consistent across dc and rf channels. Averaged over both channels, the linear coefficient $a_1$ is reduced by approximately $1.2$–$1.3$~dB for squeezed probing and increased by approximately $2.3$–$2.9$~dB for antisqueezed probing relative to coherent light, depending weakly on probe power and ensemble polarization. Independent constant fits of $\mathcal{S}^{\mathrm{PSN}}$ as a function of pump power at fixed probe power  further confirm the absence of pump-power dependence and yield consistent dB-level noise suppression and enhancement factors.

\PRLsection{SPN}
At low probe powers, the SPN exhibits a quadratic dependence on the probe power, $\mathcal{S}^{\mathrm{SPN}}\propto P\subpr^{2}$, and shows no measurable dependence on the pump power nor on the quantum state of the probe field. This behavior is consistent with intrinsic quantum fluctuations of the collective spin that are transferred to the optical readout with a gain proportional to the probe intensity. At higher probe powers, the quadratic scaling of the SPN power spectral density breaks down and is accompanied by a broadening of the spin-noise linewidth, as shown in the \SM. This deviation is attributed to scattering of the probe light, which introduces an additional relaxation channel proportional to the probe power. As a result, a fixed total spin-noise power is redistributed over an increasing bandwidth, leading to a reduced peak noise spectral density while preserving the overall quadratic scaling of the integrated spin-noise power. Quadratic fits to the total SPN power are shown in Fig.~\ref{fig:allNoises} (center). The extracted quadratic coefficients are shown in the \SM~and are consistent across dc and rf channels and show only weak variations with the probe quantum state, confirming that SPN is independent of optical squeezing. 

\PRLsection{MBA} Fig.~\ref{fig:allNoises} (right) shows the MBA total noise power scaling, where
fits of the form $\mathcal{S}^{\mathrm{MBA}} = a_0 + a_3 P\subpr^{3} $ yield cubic coefficients $a_3$ that increase strongly with pump power, consistent with the expected scaling of back-action arising from probe-induced optical Zeeman shifts. Independent fits to the pump-power dependence at fixed probe power confirm this behavior and yield consistent quadratic coefficients for both measurement channels. A small offset term $a_0$ is observed in the cubic fits, which increases with pump power but remains subdominant over the explored probe-power range. We attribute this offset to technical background contributions rather than to a fundamental back-action process. It does not affect the extracted scaling behavior.\\

Taken together, these results reveal a resource-dependent optimization landscape intrinsic to continuously operating multiparameter quantum sensors. While increasing probe power suppresses photon shot noise, it simultaneously enhances spin-noise transduction and, more critically, measurement back-action, leading to a nonmonotonic dependence of sensitivity on probe intensity. In contrast to back-action-evading single-parameter sensors, where increased measurement strength is unconditionally beneficial \cite{troullinou2021squeezed,TroullinouPRL2023}, multiparameter operation of the hOPM imposes a fundamental trade-off between high-frequency sensitivity, limited by PSN, and low-frequency sensitivity, limited by MBA. The observed cubic scaling of MBA with probe power and its quadratic dependence on pump power therefore imply the existence of an optimal operating point jointly determined by both optical resources. The origin and nature of this optimum are not specific to the present implementation but are expected to be found broadly in continuously monitored spin systems.

In conclusion, we have experimentally established the scaling laws governing photon shot noise, spin projection noise, and measurement back-action noise in a continuously operating spin-precession sensor, with and without quantum enhancement in the form of squeezed probe light. Using a paradigmatic multi-parameter sensor, the hybrid rf–dc optically pumped magnetometer, we observe linear, quadratic, and cubic scaling with probe power of PSN, SPN, and MBA, respectively, together with a quadratic dependence of MBA on pump power. The measured scalings are in agreement with a stochastic Bloch-equation model that captures both the spin dynamics and the quantum properties of the optical readout. The same scalings are observed for rf and dc measurements, and  apply to a wide variety of sensors that employ continuous optical pumping and non-destructive readout of spin ensembles.  These results, together with earlier results on scaling of quantum noise with the size of the spin ensemble \cite{TroullinouPRL2023}, provide a full picture of the quantum noise landscape for such sensors, and a guide to their design and optimization. \\

\paragraph*{Acknowledgments}
The authors thank Theo Meireles for useful discussions and Piotr Sierant for his help with data processing and many sarcastic remarks about plots. This work was supported by 
European Commission projects Field-SEER (ERC 101097313), OPMMEG (101099379) and QUANTIFY (101135931); Spanish Ministry of Science MCIN project SAPONARIA (PID2021-123813NB-I00) and SALVIA (PID2024-158479NB-I00),  MARICHAS (PID2021-126059OA-I00), ``NextGenerationEU/PRTR.'' (Grant FJC2021-047840-I) and ``Severo Ochoa'' Center of Excellence CEX2024-001490-S [MICIU/AEI/10.13039/501100011033];  Generalitat de Catalunya through the CERCA program,  DURSI grant No. 2021 SGR 01453 and QSENSE (GOV/51/2022).  Fundaci\'{o} Privada Cellex; Fundaci\'{o} Mir-Puig.
Funded by the European Union. DMA acknowledges funding from the European Union’s Horizon Europe research and innovation programme under the MSCA Grant Agreement No. 101081441. Views and opinions expressed are however those of the author(s) only
and do not necessarily reflect those of the European Union. Neither
the European Union nor the granting authority can be held responsible for them. \\

\newpage
\pagebreak
\newpage 

\onecolumngrid 
\section*{End Matter}

\twocolumngrid 

\paragraph{Experimental setup and operating parameters}
\label{sec:experiment}
The basic construction and operating principles of the hOPM, as well as the generation of polarization-squeezed probe light, have been described in detail previously \cite{lipka2024multiparameter,Predojevic2008,troullinou2021squeezed, SqhOPM2025}. Here we summarize the key elements relevant to the quantum-noise scaling measurements presented in this article. The dc magnetic field $\bvdc$ is applied in the $x$--$z$ plane at $45^\circ$ with respect to the optical pumping and probing direction, enabling simultaneous sensitivity to dc and rf magnetic fields. A weak rf magnetic field $\bvrf$ is applied along the $x$ axis and oscillates at frequency $\omega_{\mathrm{rf}}$, which is phase synchronized with the optical pumping frequency $\omega_{\mathrm{p}}$. Optical pumping is implemented using the Bell--Bloom technique \cite{bell1961optically}, in which the collective atomic spin $\mathbf{F}$ is driven periodically at $\omega_{\mathrm{p}} \approx \omega_L$, producing a resonant buildup of transverse spin polarization. Isotopically enriched $^{87}$Rb vapor is contained in a magnetically shielded glass cell of internal length \SI{3}{\centi\meter}, filled with \SI{100}{\torr} of N$_2$ buffer gas. The cell is heated to \SI{105}{\celsius}, corresponding to an atomic number density of approximately \SI{8.2e12}{atoms\per\centi\meter\cubed}. Magnetic fields are generated by low-noise induction coils driven by precision current sources, producing a dc bias field of $\bvdc \approx \SI{6}{\micro\tesla}$ ($\omega_L \approx 2\pi \times \SI{42}{\kilo\hertz}$) and the rf drive field. Optical pumping is provided by a circularly polarized beam propagating along the $z$ axis, with instantaneous pump power varied between \SI{5}{\micro\watt} and \SI{15}{\micro\watt}. The pump laser frequency is modulated at $\omega_{\mathrm{p}}$ around the $^{87}$Rb D$_1$ transition, sweeping through resonance twice per Larmor precession cycle. The modulation signal is phase and frequency synchronized with the rf drive. The atomic spin precession is monitored via off-resonant Faraday rotation of a linearly polarized probe beam. The probe power is varied between \SI{0.5}{\milli\watt} and \SI{3}{\milli\watt} to study the scaling of the quantum noise contributions. The probe light is derived from a frequency-stabilized diode laser system operating at \SI{795}{\nano\meter}, blue-detuned by approximately \SI{50}{\giga\hertz} from the D$_1$ transition. Polarization rotation is detected using a shot-noise-limited balanced polarimeter consisting of a half-wave plate, a Wollaston prism, and a balanced photodetector. Polarization-squeezed probe states are generated by combining a vertically polarized squeezed vacuum field, produced by a subthreshold optical parametric oscillator, with a horizontally polarized local oscillator. The relative phase between the squeezed vacuum and the local oscillator is actively stabilized. The measured polarization squeezing is approximately 2.0~dB before the atomic cell and 1.8~dB after the cell at the balanced detector. Detector outputs and the pump modulation signal are digitized for digital demodulation and spectral analysis. In the present experiment, the achievable squeezing is limited by optical losses, residual phase noise, and the available pump power for squeezed-light generation.

\begin{figure}[t]
    \centering
    \includegraphics[width=1\columnwidth]{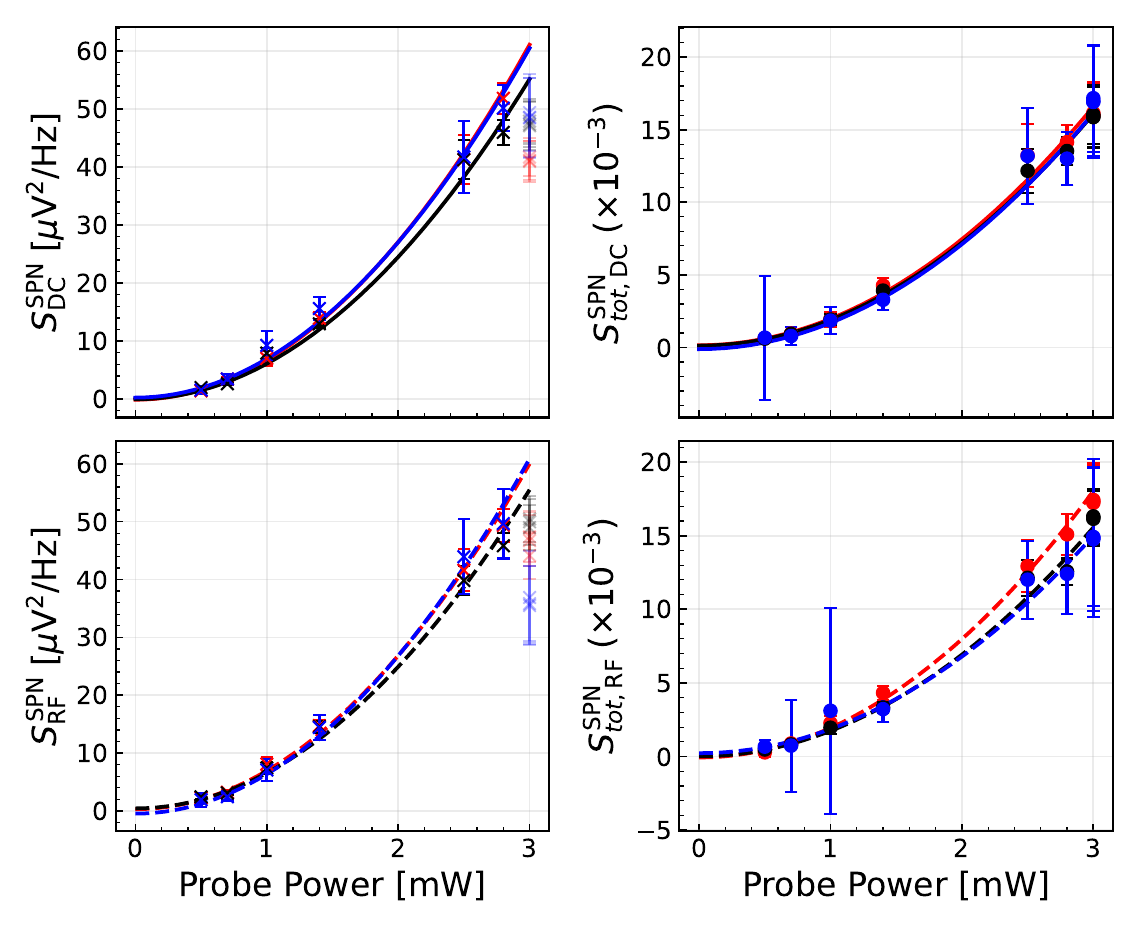}
\caption{SPN power spectral density and total SPN power as a function of probe power for an unpolarized atomic ensemble, probed with coherent (red), squeezed (black), and antisqueezed (blue) light, shown for both dc and rf measurement channels. Dashed lines indicate quadratic fits of the form $\mathcal{S}^{\mathrm{SPN}} = a_2 P_{\mathrm{pr}}^{2} + a_0$ to the probe-power dependence; the corresponding fit parameters are summarized in Table~\ref{tab:spn_combined_fits}. The data point at $P_{\mathrm{pr}}=3$~mW is excluded from the quadratic fit to $\mathcal{S}^{\mathrm{SPN}}$ due to probe-induced linewidth broadening but remains consistent with the quadratic scaling of the total SPN noise power.}
    \label{fig:SPN}
\end{figure}

\paragraph{Noise scaling and fit parameters}
Figure~\ref{fig:SPN} summarizes the measured scaling of SPN contribution to the polarization noise for both the dc and rf readout channels. 

\begin{table*}[t]
\centering
\caption{
Linear fits to the PSN power
$\mathcal{S}^{\mathrm{PSN}} = a_1 P_{\mathrm{pr}} + a_0$
as a function of probe power for dc and rf channels.
Results are shown for coherent (coh), squeezed (sq), and antisqueezed (asq)
probing, for unpolarized and polarized atomic ensembles,
and for pump powers of 5~µW and 10~µW.
Each dB column gives the relative change of the linear coefficient $a_1$
with respect to the coherent state of the same channel, ensemble, and pump power.
Uncertainties represent one standard deviation.
}
\label{tab:psn_probe_db}

\setlength{\tabcolsep}{3pt}
\renewcommand{\arraystretch}{1.15}

\resizebox{1\linewidth}{!}{%
\begin{tabular}{c ccc ccc ccc ccc}
\hline\hline
& \multicolumn{12}{c}{$\mathcal{S}^{\mathrm{PSN}}$ [$\mu$V$^2$/Hz]} \\
\cline{2-13}
State
& \multicolumn{6}{c}{unpolarized}
& \multicolumn{6}{c}{polarized} \\
\cline{2-7}\cline{8-13}
& \multicolumn{3}{c}{5 µW\footnote{Pump light was blocked for this measurement.}}
& \multicolumn{3}{c}{10 µW$^\mathrm{a}$}
& \multicolumn{3}{c}{5 µW}
& \multicolumn{3}{c}{10 µW} \\
\cline{2-13}
& $a_1$ & $a_0$ & dB
& $a_1$ & $a_0$ & dB
& $a_1$ & $a_0$ & dB
& $a_1$ & $a_0$ & dB \\
\hline

$\mathrm{coh}_{\mathrm{dc}}$
& $19.59(15)$ & $-0.48(11)$ & $0$
& $19.25(6)$ & $-1.77(6)$ & $0$
& $19.65(15)$ & $-0.42(11)$ & $0$
& $19.15(5)$ & $-1.66(6)$ & $0$ \\

$\mathrm{sq}_{\mathrm{dc}}$
& $14.50(12)$ & $-0.60(9)$ & $-1.31(5)$
& $14.51(7)$ & $-2.04(10)$ & $-1.31(4)$
& $14.65(12)$ & $-0.68(9)$ & $-1.30(5)$
& $14.90(5)$ & $-1.95(5)$ & $-1.11(3)$ \\

$\mathrm{asq}_{\mathrm{dc}}$
& $36.25(69)$ & $9.25(71)$ & $+2.67(6)$
& $37.30(14)$ & $3.90(15)$ & $+2.88(3)$
& $34.74(36)$ & $8.46(39)$ & $+2.47(4)$
& $33.15(15)$ & $7.03(15)$ & $+2.38(4)$ \\

\hline

$\mathrm{coh}_{\mathrm{rf}}$
& $19.60(15)$ & $-0.43(11)$ & $0$
& $19.57(6)$ & $-1.99(6)$ & $0$
& $19.66(14)$ & $-0.49(10)$ & $0$
& $19.44(6)$ & $-1.89(6)$ & $0$ \\

$\mathrm{sq}_{\mathrm{rf}}$
& $14.72(12)$ & $-0.71(9)$ & $-1.24(6)$
& $15.64(10)$ & $-3.09(15)$ & $-0.98(5)$
& $14.87(12)$ & $-0.79(9)$ & $-1.20(6)$
& $14.87(6)$ & $-1.87(5)$ & $-1.17(4)$ \\

$\mathrm{asq}_{\mathrm{rf}}$
& $37.85(64)$ & $5.53(58)$ & $+2.88(7)$
& $34.74(19)$ & $4.67(27)$ & $+2.50(4)$
& $34.79(35)$ & $8.14(37)$ & $+2.69(4)$
& $31.53(15)$ & $7.54(16)$ & $+2.21(4)$ \\

\hline\hline
\end{tabular}%
}
\end{table*}

\begin{table}[t]
\centering
\caption{
PSN power $\mathcal{S}^{\mathrm{PSN}}$ at a fixed probe power
$P_{\mathrm{pr}} = 3~\mathrm{mW}$, obtained from constant fits as a function
of pump power for dc and rf channels. Results are shown for coherent (coh), squeezed (sq), and antisqueezed (asq)
probing, for unpolarized and polarized atomic ensembles.
The dB columns give the relative change with respect to the coherent state of the same channel and ensemble.
Uncertainties represent one standard deviation.
}
\label{tab:psn_pump}

\begin{tabular}{c ccc ccc}
\hline\hline
& \multicolumn{6}{c}{$\mathcal{S}^{\mathrm{PSN}}$ [$\mu$V$^2$/Hz]} \\
\cline{2-7}
State
& \multicolumn{3}{c}{unpolarized}
& \multicolumn{3}{c}{polarized} \\
\cline{2-4}\cline{5-7}
& $a_0$ & dB
& 
& $a_0$ & dB
& \\
\hline
$\mathrm{coh}_{\mathrm{dc}}$
& $55.11(16)$ & $0$
&
& $55.17(20)$ & $0$
& \\

$\mathrm{sq}_{\mathrm{dc}}$
& $42.17(14)$ & $-1.15(1)$
&
& $41.93(16)$ & $-1.19(1)$
& \\

$\mathrm{asq}_{\mathrm{dc}}$
& $134.00(40)$ & $+3.86(1)$
&
& $121.88(44)$ & $+3.44(2)$
& \\
\hline
$\mathrm{coh}_{\mathrm{rf}}$
& $54.80(16)$ & $0$
&
& $54.99(20)$ & $0$
& \\

$\mathrm{sq}_{\mathrm{rf}}$
& $42.04(16)$ & $-1.15(2)$
&
& $41.90(17)$ & $-1.19(2)$
& \\

$\mathrm{asq}_{\mathrm{rf}}$
& $134.70(41)$ & $+3.90(1)$
&
& $122.52(41)$ & $+3.48(2)$
& \\
\hline\hline
\end{tabular}%
\end{table}

\begin{table}[t]
\centering
\caption{Quadratic fits to the SPN power spectral density
$\mathcal{S}^{\mathrm{SPN}} = a_0 + a_2 P_{\mathrm{pr}}^2 $ and to the total
SPN power
$\mathcal{S}^{\mathrm{SPN}}_{\mathrm{tot}} = a_0 + a_2 P_{\mathrm{pr}}^2$.
Total-noise coefficients include the $\pi/2$ Lorentzian integration factor.
The dB columns give the relative change of the quadratic coefficient $a_2$
with respect to the coherent state of the same measurement channel.}
\label{tab:spn_combined_fits}

\resizebox{1\linewidth}{!}{%
\begin{tabular}{c ccc ccc}
\hline\hline
& \multicolumn{3}{c}{$\mathcal{S}^{\mathrm{SPN}}$ [$\mu$V$^2$/Hz]}
& \multicolumn{3}{c}{$\mathcal{S}^{\mathrm{SPN}}_{\mathrm{tot}}$ [$\mu$V$^2$]} \\
\cline{2-4}\cline{5-7}
State
& $a_2$ & $a_0$ & dB
& $a_2$ & $a_0$ & dB \\
\hline
$\mathrm{coh}_{\mathrm{dc}}$
& $6.74(2)$ & $-0.07(2)$ & $0$
& $1850(80)$ & $143(63)$ & $0$ \\

$\mathrm{sq}_{\mathrm{dc}}$
& $6.28(2)$ & $+0.06(2)$ & $-0.31(1)$
& $1800(70)$ & $121(46)$ & $-0.12(2)$ \\

$\mathrm{asq}_{\mathrm{dc}}$
& $6.48(4)$ & $+0.3(5)$ & $-0.17(3)$
& $1860(220)$ & $-83(410)$ & $+0.01(5)$ \\
\hline
$\mathrm{coh}_{\mathrm{rf}}$
& $6.75(3)$ & $-0.04(2)$ & $0$
& $2030(90)$ & $-83(99)$ & $0$ \\

$\mathrm{sq}_{\mathrm{rf}}$
& $6.27(2)$ & $-0.06(2)$ & $-0.31(2)$
& $1765(65)$ & $17(88)$ & $-0.60(2)$ \\

$\mathrm{asq}_{\mathrm{rf}}$
& $6.91(6)$ & $-0.7(5)$ & $+0.11(4)$
& $1690(280)$ & $162(640)$ & $-0.80(7)$ \\
\hline\hline
\end{tabular}%
}
\end{table}

\begin{table}[t]
\centering
\caption{
Cubic fits to the total MBA noise
$\mathcal{S}^{\mathrm{MBA}}_{\mathrm{tot}} = a_3 P_{\mathrm{pr}}^{3} + a_0$.
All coefficients include the $\pi/2$ Lorentzian integration factor.
Uncertainties represent one standard deviation.
The dB columns give the relative change of the cubic coefficient $a_3$
with respect to the coherent state of the same channel and pump power.
}
\label{tab:mba_probe_sep_pump}

\resizebox{1\linewidth}{!}{%
\begin{tabular}{c cccc cccc}
\hline\hline
& \multicolumn{8}{c}{$\mathcal{S}^{\mathrm{MBA}}_{\mathrm{tot}}$ [$\mu$V$^2$]} \\
\cline{2-9}
State
& \multicolumn{4}{c}{5 µW}
& \multicolumn{4}{c}{10 µW} \\
\cline{2-5}\cline{6-9}
& $a_3$ & $a_0$ & dB & 
& $a_3$ & $a_0$ & dB &  \\
\hline
$\mathrm{coh}_{\mathrm{dc}}$
& $6.0(9.3)$ & $120(260)$ & $0$
&
& $320(68)$ & $1520(260)$ & $0$ \\

$\mathrm{sq}_{\mathrm{dc}}$
& $39(77)$ & $36(137)$ & $+8.1(9)$
&
& $215(53)$ & $1360(255)$ & $-1.7(4)$ \\

$\mathrm{asq}_{\mathrm{dc}}$
& $-138(190)$ & $2090(1740)$ & ---
&
& $298(105)$ & $1490(650)$ & $-0.3(5)$ \\
\hline
$\mathrm{coh}_{\mathrm{rf}}$
& $-33(99)$ & $492(335)$ & $0$
&
& $276(75)$ & $1660(250)$ & $0$ \\

$\mathrm{sq}_{\mathrm{rf}}$
& $16(77)$ & $357(231)$ & ---
&
& $291(57)$ & $1230(190)$ & $+0.2(3)$ \\

$\mathrm{asq}_{\mathrm{rf}}$
& $5.8(168)$ & $30(750)$ & ---
&
& $239(145)$ & $2750(1240)$ & $-0.6(7)$ \\
\hline\hline
\end{tabular}%
}
\end{table}

\begin{table}[t]
\centering
\caption{
Quadratic scaling coefficients for total MBA noise
$\mathcal{S}^{\mathrm{MBA}}_{\mathrm{tot}} = a_2\,P_{\mathrm{pump}}^{2} + a_0$.
All coefficients include the $\pi/2$ Lorentzian integration factor.
The dB column gives the relative change of the quadratic coefficient $a_2$
with respect to the coherent state of the same measurement channel.
}
\label{tab:mba_pump_quadratic_db}

\setlength{\tabcolsep}{4pt}
\renewcommand{\arraystretch}{1.15}
\begin{tabular}{c ccc}
\hline\hline
& \multicolumn{3}{c}{$\mathcal{S}^{\mathrm{MBA}}_{\mathrm{tot}}$ [$\mu$V$^2$]} \\
\cline{2-4}
State & $a_2$ & $a_0$ & dB \\
\hline
$\mathrm{coh}_{\mathrm{dc}}$
& $85.8(24)$ & $-1340(2950)$ & $0$ \\

$\mathrm{sq}_{\mathrm{dc}}$
& $82.5(20)$ & $-1070(2640)$ & $-0.17(12)$ \\

$\mathrm{asq}_{\mathrm{dc}}$
& $72.8(38)$ & $-1550(5140)$ & $-0.72(23)$ \\
\hline
$\mathrm{coh}_{\mathrm{rf}}$
& $88.3(27)$ & $-2400(3280)$ & $0$ \\

$\mathrm{sq}_{\mathrm{rf}}$
& $86.6(22)$ & $-1180(2590)$ & $-0.08(11)$ \\

$\mathrm{asq}_{\mathrm{rf}}$
& $67.3(44)$ & $454(6020)$ & $-1.19(29)$ \\
\hline\hline
\end{tabular}
\end{table}

\noindent
At low probe powers, the SPN exhibits a quadratic dependence on the probe power, $\mathcal{S}^{\mathrm{SPN}}\propto P_{\mathrm{pr}}^{2}$, and shows no measurable dependence on the pump power or on the quantum state of the probe field. At higher probe powers, the quadratic scaling of the SPN power spectral density breaks down and is accompanied by a broadening of the spin-noise linewidth. This deviation is attributed to probe-induced optical pumping, which introduces an additional relaxation channel proportional to the probe power. As a result, a fixed total spin-noise power is redistributed over an increasing bandwidth, leading to a reduced peak noise spectral density while preserving the overall quadratic scaling of the integrated spin-noise power. Quadratic fits to the SPN power spectral density and to the total SPN power are summarized in Table~\ref{tab:spn_combined_fits}. The data point at the highest probe power ($P_{\mathrm{pr}}=3$~mW) is excluded from the quadratic fit to $\mathcal{S}^{\mathrm{SPN}}$ due to the increased linewidth but remains consistent with the expected quadratic scaling of the total SPN power.
The fit parameters for all fits shown in Fig.~2 are listed in Tables~\ref{tab:psn_probe_db}–\ref{tab:mba_pump_quadratic_db}.

\newpage
\pagebreak
\newpage 

\onecolumngrid 
\twocolumngrid 

\newpage
\pagebreak
\newpage 

\onecolumngrid 
\twocolumngrid 

%

\end{document}